\documentclass[preprint,showpacs,preprintnumbers,amsmath,amssymb]{revtex4}
\usepackage{amsmath,graphicx,amssymb}
\newcommand{\nnn}{\nonumber \\}

\def\nn{\nonumber}

\def\beq{\begin{equation}}
\def\eeq{\end{equation}}
\def\bea{\begin{eqnarray}}
\def\eea{\end{eqnarray}}

\def\Tr{\rm Tr}

\newcommand{\mt}{\mathcal}
\newcommand{\tl}{\tilde}
\begin{document}
\preprint{HUPD 0904}
\pacs{11.30.Qc, 11.30.Pb, 12.60.Jv}
\title
{Threshold corrections to
the radiative breaking of electroweak symmetry
and neutralino dark matter\\
in supersymmetric seesaw model}
\author{{\bf
Sin Kyu Kang $^{\rm a}$, Akina Kato$^{\rm b}$, Takuya Morozumi$^{\rm b}$
 and Norimi Yokozaki$^{\rm b}$}\\
$^{\rm a}$ School of Liberal Arts, Seoul National University of Technology, Seoul 139-473, Korea.\\
$^{\rm b}$ Graduate School of Science, Hiroshima University,
Higashi-Hiroshima, 739-8526, Japan}
%\date{}

\begin{abstract}
We study the radiative electroweak symmetry breaking and the relic abundance of neutralino dark matter in supersymmetric type I seesaw model.
In this model, there exist threshold corrections to Higgs bilinear terms coming from heavy singlet sneutrino loops, which
make the soft supersymmetry breaking (SSB) mass for up-type Higgs shift at the seesaw scale and
thus a minimization condition for the Higgs potential is affected.
We show that the required fine-tuning between the Higgsino mass parameter $\mu$ and
SSB mass for up-type Higgs may be reduced at electroweak scale,
 due to the threshold corrections. We also present how the parameter $\mu$ depends on SSB B-parameter for  heavy singlet sneutrinos.
Since the property of neutralino dark matter is quite sensitive to the size of $\mu$,
we discuss how the relic abundance of neutralino dark matter is affected by the SSB B-parameter.
Taking the SSB B-parameter of order of a few hundreds TeV, the required relic abundance of
neutralino dark matter can be correctly achieved. In this case, dark matter is a mixture of bino and Higgsino, under the condition that
gaugino masses are universal at the grand unification scale.

\end{abstract}
\maketitle

\section{Introduction}

Supersymmetric (SUSY) seesaw model is a SUSY extension of seesaw model \cite{seesaw_yanagida, seesaw_gellman} which naturally explains small masses of neutrinos and stabilizes the hierarchy between electroweak scale and some other high scale without severe fine-tuning if the mass spectrum of
super-partners are less than TeV scale as well.
In  SUSY type I seesaw model, we introduce not only heavy right-handed(RH) Majorana neutrinos but also their super partner called sneutrinos
which are standard model gauge singlet.
%In type I SUSY seesaw model, super partner of the heavy Majorana
%neutrinos which are absent in the minimal SUSY standard model (MSSM) are also introduced.
This leads us to anticipate that some predictions of MSSM can be deviated due to the contributions associated with the heavy RH neutrinos and their super partners, 
and new phenomena absent in MSSM may occur in  SUSY type I seesaw model.
With this regards, there have been attempts to study lepton flavor violation
and neutrino masses in SUSY type I seesaw model\cite{seesaw_flavor1, seesaw_flavor2, susyseesaw_hisano}.
On the other hand, the gauge singlet RH neutrino superfield may affect Higgs sector as investigated in Ref. \cite{seesaw_split}, where
they have shown that there is a sizable negative loop contribution to the mass of the lightest Higgs field in the split-SUSY scenario
at the price of giving up the naturalness in supersymmetry.

%However in the split-SUSY scenario, typical scale of soft scalar masses
%are much higher than the electroweak scale. Therefore if we consider the supersymmetry as a solution to naturalness
%problem seriously, split-SUSY scenario is not preferable.
In this study, we revisit the issue as to how the Higgs sector can be affected by heavy singlet sneutrinos while keeping naturalness in supersymmetry.
It is well known that the lightest CP-even Higgs mass in the MSSM  can get large one-loop corrections which increase with the top quark and squark
masses \cite{Okada:1990vk,Okada:1990gg,Haber:1990aw,Ellis:1990nz}.
The current experimental bound on the lightest CP-even Higgs mass, $m_h\gtrsim 114$ GeV, demands top squark mass to be larger than 500 GeV
\cite{Kobayashi:2006fh},
which in turn leads to a fairly large correction to the soft supersymmetry breaking (SSB) mass for the up-type Higgs $m_{H_2}^2$.
In the MSSM, electroweak symmetry can be broken due to the large logarithmic correction to $m_{H_2}^2$ \cite{Inoue:1982ej,Inoue:1982pi,Inoue:1983pp,Ibanez:1982fr,AlvarezGaume:1981wy}.
However, as is known, we need rather large fine-tuning between the Higgsino mass parameter $\mu$ and SSB mass $m_{H_2}^2$
to achieve the $Z$-boson mass at the electroweak scale through a minimization condition for the Higgs potential of the MSSM .
In this study, we show that there exist some new contributions generated from the loops mediated by heavy singlet sneutrino sector to SSB mass $m_{H_2}^2$ and the Higgsino mass parameter $\mu$ in SUSY type I seesaw model. The new contributions are given in terms of SSB parameters $B_N$ and SSB mass term for the singlet sneutrino $m^2_{\tilde{N}}$
at the seesaw scale.

Integrating out the singlet neutrino superfield below the seesaw scale,
SUSY type I seesaw becomes equivalent to the MSSM but those new contributions are taken to be
threshold corrections to Higgs bilinear terms. 
As will be discussed, those threshold corrections can lower the sizes of $m_{H_2}^2$ and $\mu$ at the electroweak scale and thus the fine-tuning may be reduced.
This means that the fine-tuning required for the radiative electroweak symmetry breaking
can be shifted to tuning the size of $B_N$ at the seesaw scale.
In this paper, we investigate how the sizes of $m_{H_2}^2$ and $\mu$ at the electroweak scale
depend on the parameter $B_N$.

Since the property of neutralino dark matter is quite sensitive to the size of $\mu$,
we discuss how the relic abundance of neutralino dark matter is affected by the parameter $B_N$.
In fact, there exist some literatures in which the impacts of neutrino Yukawa couplings on neutralino dark matter in SUSY type I seesaw model
have been discussed \cite{Barger,Calibbi,seesaw_DM1,seesaw_DM2,seesaw_DM3,seesaw_DM4,seesaw_DM5,seesaw_DM6}. 
It has been found that some regions of parameter space  can significantly affect the  neutralino relic density without the threshold corrections associated with the heavy singlet neutrino superfield.
In our work, however, we consider possible existence of the threshold corrections generated from the loops mediated via the heavy singlet neutrino superfield which can also significantly affect the neutralino relic abundance by lowering the sizes of $m_{H_2}^2$ and $\mu$ at the electroweak scale.
Such a possibility of the impact on the neutralino relic density has not been studied before.

This paper is organized as follows.
First, we present the effective potential for Higgs fields in SUSY type I  seesaw model in section II.
We show that  threshold corrections to Higgs bilinear terms are generated from the loops
mediated by heavy singlet neutrino superfields. 
In section III, we give the alternative derivation for the threshold corrections, using renormalization group equations (RGEs) for a general field
theory.
In section IV, we study the contributions of the threshold corrections to the radiative electroweak symmetry breaking and investigate
how  the size of the parameter $\mu$ can be affected by them.
In section V, we discuss the relic abundance of neutralino dark matter.
Finally section VI is devoted to conclusions and discussions.
The details of convention for CP phases and derivation of the effective potential for Higgs fields are given in Appendix.

\section{The Effective Potential of SUSY Type I Seesaw Model}
In this section, we first derive the effective potential of SUSY type I seesaw model, and then show that
there exist threshold corrections to Higgs bilinear terms arisen due to the heavy RH singlet sneutrinos.
Those threshold corrections may be modified by wave function renormalization for Higgs field.
%Therefore we comment on the wave function renormalization from neutrinos and sneutrinos.

The super potential of the SUSY seesaw model is given by
\begin{eqnarray}
 W = \mu H_1 \cdot H_2 - Y_\nu (\hat{L} \cdot \hat{H}_2) \hat{N}^c -
\frac{M_R}{2} \hat{N}^c \hat{N}^c, \label{superpotential}
\end{eqnarray}
where $\hat{N}^c$ is a gauge singlet chiral superfield, which contains a RH neutrino and its scalar partner.
$M_R$ denotes the mass of the RH neutrino.
Here, we do not consider the terms associated with the charged leptons and quarks whose contributions to our study are negligibly
small except for top quark superfield. From now on, we consider only one generation of $\hat{N}^c$ for simplicity, and the extension to three generations is straight-forward.
The soft breaking terms of the Lagrangian in SUSY seesaw model are given by
\begin{eqnarray}
{\cal L}_{\rm soft}&=&
- m_{\tilde{L}}^2 |\tilde{L}|^2- m_{\tilde{N}}^2 |\tilde{N}|^2
-\left(\frac{1}{2} B_N^{\ast} M_R^{\ast} \tilde{N}^2+ h.c.
\right) \nnn
&&+  2 {\rm Re}(B \mu H_1 \cdot H_2) -m_{H_1}^2 H_1^{\dagger}  H_1
-m_{H_2}^2 H_2^{\dagger}H_2 \nnn
&& +\left(A_\nu Y_\nu (H_2 \cdot \tilde{L}) N^* + h.c. \right) ,
\label{softbr}
\end{eqnarray}
where we can take $M_R, B_N, Y_\nu, \mu$ to be real by superfield rotation and $U(1)_R$ symmetry, whereas $A_\nu$ and $B$ are left as complex numbers.
We discuss the details of the phase convention in Appendix A.
From the superpotential given in Eq. (\ref{superpotential}), the SUSY part of the Lagrangian
is obtained as follows:
\bea
{\cal L}_{\rm susy}&=&-|Y_{\nu} \tilde{L} \cdot H_2+ M_R \tilde{N}^{\ast}|^2
-|Y_{\nu} \tilde{N}^{\ast} \tilde{L}- \mu H_1|^2-|\mu|^2 H_2^{\dagger} H_2
-Y_{\nu}^2 |\tilde{N}|^2 H_2^{\dagger} H_2\nn \\
 &-& \frac{1}{2} M_R \overline{N_R} {N_R}^c -Y_{\nu}
\overline{N_R} l_L \cdot H_2 + {\rm h.c.}.
\eea

With this Lagrangian, we can derive the effective potential by using field dependent masses for the singlet RH neutrinos and sneutrinos.
The effective Higgs potential which includes 1-loop contributions mediated by the singlet RH neutrino superfields is written as
\bea
V_{eff}^{\rm 1 loop}
&=&  \left(|\mu|^2 + m_{H_1}^2(Q^2) \right)  H_1^{\dagger}H_1
+\left(|\mu|^2 + m_{H_2}^2(Q^2) \right)  H_2^{\dagger}H_2
-2 {\rm Re.} (B(Q^2) \mu H_1 \cdot {H}_2)
\nn \\
&+&  \left(\mu^2 \frac{Y_\nu^2}{16 \pi^2}
\log \frac{M_R^2}{Q^2} \right) H_1^{\dagger} H_1 \nn \\
&+&  \frac{Y_{\nu}^2}{16 \pi^2}
\left(
\log \frac{M_R^2}{Q^2} (m_{\tilde L}^2+m_{\tilde N}^2
+|A_\nu^2|)
+2 m_{\tilde N}^2 + 2 {\rm Re.}(A_\nu B_N)
\right)H_2^{\dagger}H_2 \nn \\
&-&2{\rm Re.} \left( \frac{Y_\nu^2}{16 \pi^2}
(B_N + A_\nu \log \frac{M_R^2}{Q^2}) \mu
H_1 \cdot H_2 \right) - \mathcal{L}_D ,
\label{sseesawhiggs}
\eea
where $Q$ is a renormalization scale and $\mathcal{L}_D$ is D-term contributions given by
\begin{eqnarray}
 \mathcal{L}_D = - \frac{g'^2}{8}\left(H_1^\dagger H_1- H_2^\dagger H_2 \right)^2
- \frac{g^2}{8}\left(H_1^\dagger \tau^a H_1 + H_2^\dagger \tau^a H_2 \right)^2 .
\end{eqnarray}
In Appendix B, we present in detail how the effective potential is derived.
Matching this effective potential with that of MSSM at the seesaw scale, we can obtain some relations between MSSM parameters and corresponding ones in SUSY seesaw model.
Here, we do not include the loop contributions mediated by top quark and its super partner because they are identical to each other in both MSSM and SUSY seesaw model, and thus canceled in the relations. Therefore
those contributions are irrelevant to the threshold corrections for the Higgs bilinear terms.
The Higgs potential of the MSSM is given by,
\bea
{V}_{\rm MSSM}&=& \left(|\mu|^2 + \bar{m}_{H_1}^2(Q^2)\right) {H_1^Q}^\dagger H_1^Q
            + \left(|\mu|^2 + \bar{m}_{H_2}^2(Q^2)\right) {H_2^Q}^\dagger H_2^Q \nn \\
&-& \left(\bar{B}(Q^2) \mu H_1^Q \cdot {H^Q_2} + h.c. \right) - \mathcal{L}_D.
\label{mssmhiggs}
\eea
By matching the Higgs potentials Eq.(\ref{mssmhiggs}) with Eq.(\ref{sseesawhiggs}) at $Q^2=M_R^2$, we obtain the following relations,
\bea
\bar{m}_{H_1}^2(M_R^2)&=&m_{H_1}^2(M_R^2) , \nn \\
\bar{m}_{H_2}^2(M_R^2)&=&m_{H_2}^2(M_R^2)+ \frac{Y_\nu^2}{8\pi^2}\left( m_{\tilde N}^2 +
  {\rm Re.}(A_\nu B_N) \right) , \nn \\
\bar{B}(M_R^2)&=&
B(M_R^2) + \frac{Y_\nu^2}{16 \pi^2}B_N . \label{eq:threshold}
\eea
%The results are consistent with our previous result of the effective lagrangian \cite{higgsseesaw_pre} calculated with cutoff regularization.
On the other hand, the wave function renormalization for the Higgs field $H_2$ in the limit of small external momenta is given by
\begin{eqnarray}
\left(1 - \frac{Y_\nu^2}{16\pi^2} \log \frac{M_R^2}{Q^2} \right) \partial_\mu H_2^{Q \dagger} \partial^\mu H_2^Q, \label{eq:wav}
\end{eqnarray}
where we neglect the terms suppressed by $M_R^{-2}$. We notice that there exist no contributions from heavy RH neutrino superfields to wave function renormalization for $H_1$.
At $Q^2 = M_R^2$,  Eq.(\ref{eq:wav}) becomes $\partial_\mu H_2^\dagger \partial^\mu H_2$, so the relations given in Eq.(\ref{eq:threshold}) are not modified by wave function renormalization.

It is worth noting that the soft breaking parameter of singlet sneutrino,
$B_N$, contributes to the Higgs mass $\bar{m}_{H_2}^2(M_R^2)$ and the parameter $B$.
We use RGEs for the soft breaking parameters of the MSSM to obtain their low energy values
below the seesaw scale $M_R$, whereas the corresponding RGEs given in the SUSY seesaw model are used above the seesaw scale.
Thus, the values of the parameters in the RH side of Eq.(\ref{eq:threshold}), $m_{H_1}^2(Q^2=M_R^2)$ and
$m_{H_2}^2(Q^2=M_R^2)$, depend on the boundary condition at further high energy scale, such as $M_{GUT}$ or $M_{\rm planck}$.
%In our numerical analysis, we assume the universal soft breaking parameters at the GUT scale for simplicity.

\section{The threshold corrections from Renormalization Group Equations}
In this section, we study the alternative
derivation of the threshold corrections given in Eq.({\ref{eq:threshold}})
by using RGEs including threshold effects. The RGEs in MSSM including threshold effects are 
discussed in Refs. \cite{threshold_mssm1,threshold_mssm2,threshold_mssm3,threshold_mssm4}.
We derive the one-loop RGEs for Higgs mass squared parameters 
in the SUSY seesaw model, by using
the formulas for RGEs of dimensional parameters in general gauge field theories \cite{rge_general}. 
Then we integrate them and obtain the threshold corrections.
Here we focus on the effects from the heavy neutrino and sneutrinos.

The key point of the derivation of the threshold corrections
is to take into account three different thresholds.
One of them corresponds to the mass of RH neutrino
($M_R$), and the others correspond to the masses of the heavy sneutrinos
, i.e., the super partners of the RH neutrino. They are two real scalar
fields and their masses are deviated from $M_R$ 
due to soft SUSY breaking terms of the sneutrinos sector
, as given by
\begin{eqnarray}
 \mathcal{L}_{\rm mass}
 =- \frac{1}{2} M_{\tilde{N}_1}^2 N_1^2 -
\frac{1}{2} M_{\tilde{N}_2}^2 N_2^2 ,
\end{eqnarray}
where $N_1$ and $N_2$ are real and imaginary part of
the complex scalar field $\tilde{N}$, respectively and are
defined as,
\begin{eqnarray}
 N_1 &=& (\tilde{N} + \tilde{N}^*)/\sqrt{2} ,\ \ N_2 = (\tilde{N} - \tilde{N}^*)/(\sqrt{2}{\rm i}).
\end{eqnarray}
The masses of the $N_1$ and $N_2$ are then give by
\begin{eqnarray}
M_{\tilde{N}_1}^2 &=& m_{\tilde{N}}^2 + M_R^2 + B_N M_R , \ \ M_{\tilde{N}_2}^2 = m_{\tilde{N}}^2 + M_R^2 - B_N M_R .
\end{eqnarray}
Since $B_N$ is real positive, the hierarchy of the three
mass scales is given by
\begin{eqnarray}
M_{\tilde{N}_1}^2 > M_R^2 >M_{\tilde{N}_2}^2.
\end{eqnarray}
Then the energy scales at which $\tilde{N}_1$, $\tilde{N}_2$
and $N_R$ are
decoupled are different each other, yielding
the threshold corrections to Higgs mass squared parameters. The Higgs mass terms are given as
\begin{eqnarray}
 \mathcal{L}_{\rm Higgs}= - m_{11}^2 |H_1|^2 - m_{22}^2 |H_2|^2 - m_{12}^2 H_1 \cdot H_2 + h.c. ,
\end{eqnarray}
where,
\begin{eqnarray}
 m_{11}^2 &=& |\mu|^2 + m_{H_1}^2 ,\nnn
 m_{22}^2 &=& |\mu|^2 + m_{H_2}^2 ,\nnn
 m_{12}^2 &=& - B \mu .
\label{Higgsmass}
\end{eqnarray}

Following \cite{rge_general}, we divide all the complex
scalar fields into their real and imaginary parts, and derive
the beta functions for the Higgs mass squared parameters
by adopting 
the step functions of the renormalization 
scale ($Q$) to take into account the thresholds.  Then we obtain the threshold corrections by integrating
the beta functions with respect to the energy scale between two mass scales of the singlet sneutrinos.

At one-loop level, the beta functions for the Higgs mass parameters are given as,
\begin{eqnarray}
(4\pi)^2 \frac{d m_{11}^2}{d \ln Q} &=& Y_\nu^2 \mu^2 \left[\theta(Q^2-M_{\tilde{N}_1}^2) + \theta(Q^2-M_{\tilde{N}_2}^2)\right] , \nnn
 (4\pi)^2 \frac{d m_{12}^2}{d \ln Q} &=& Y_\nu^2 A_\nu \mu \left[\theta(Q^2-M_{\tilde{N}_1}^2) + \theta(Q^2-M_{\tilde{N}_2}^2)\right] \nnn
&& - Y_\nu^2 \mu M_R \, \theta(M_{\tilde{N}_1}^2-Q^2)\theta(Q^2 - M_{\tilde{N}_2}^2) , \nnn
 (4\pi)^2 \frac{d m_{22}^2}{d \ln Q}&=& Y_\nu^2 (m_{\tilde{N}}^2 + |A_\nu|^2)\left[\theta(Q^2-M_{\tilde{N}_1^2}^2)
+ \theta(Q^2-M_{\tilde{N}_2}^2)\right]\nnn
&& - Y_\nu^2 \left[2 {\rm Re}(A_\nu) + B_N \right] M_R \, \theta(M_{\tilde{N}_1}^2-Q^2)\theta(Q^2 - M_{\tilde{N}_2}^2) \nnn
&& + 2 Y_\nu^2 M_R^2 \left[ \theta(Q^2-M_{\tilde{N}_1}^2) + \theta(Q^2-M_{\tilde{N}_2}^2) - 2\theta(Q^2-M_R^2)\right] \nnn
&& + 2 Y_\nu^2 m_{22}^2 \, \theta(Q^2-M_R^2) + Y_\nu^2 m_{\tilde{L}}^2 \left[\theta(Q^2-M_{\tilde{N}_1}^2) + \theta(Q^2-M_{\tilde{N}_2}^2)\right] .
\label{RGEs}
\end{eqnarray}
Here, we note that only the terms coming from the
neutrino-sneutrino sector are presented because the other terms are the same as those in MSSM.
In deriving the RGEs, we take into account the
fact that the effective theory changes as passing
each threshold corresponding to the heavy degree of freedom. 
At the energy scale above $M_{\tilde{N}_1}$ where the RH
neutrino and sneutrinos are active, our RGEs given in Eq.(\ref{RGEs})
are consistent with those in supersymmetric type I
seesaw model \cite{seesaw_rge,seesaw_rge2}.
While  the
RH neutrino and the lighter sneutrino are active between the two scales $M_{\tilde{N}_1}$ and $M_R$, only the lighter sneutrino is active between the two scales $M_R$ and $M_{\tilde{N}_2}$.
Finally, the effective theory becomes MSSM below $M_{\tilde{N}_2}$.
In each step, we integrate out the heavier degrees of the freedom
and derive the effective theories which are valid at the lower energy
scales.

By integrating the beta functions with respect to $Q$
from  $M_{\tilde{N}_1}$ down to $M_{\tilde{N}_2}$,
we obtain the threshold corrections.
Since the integrals can be approximated as follows;
\begin{eqnarray}
 \int_{M_{\tilde{N}_2}}^{M_{\tilde{N}_1}} d \ln Q&=&
\ln \frac{M_{\tilde{N}_1}}{M_{\tilde{N}_2}} =
\frac{B_N}{M_R} + \mathcal{O}(M_R^{-3}), \nonumber \\
 \int_{M_{R}}^{M_{\tilde{N}_1}} d \ln Q &=&
\ln \frac{M_{\tilde{N}_1}}{M_R} = \frac{1}{2} \left[\frac{B_N}{M_R} + \frac{m_{\tilde{N}^2}}{M_R^2} - \frac{B_N^2}{2M_R^2}+
\mathcal{O}(M_R^{-3})\right].
\end{eqnarray}
Only the terms proportional to $M_R$ or $M_R^2$ in
Eq.(\ref{RGEs}) contribute to the threshold corrections. The results of integrating the beta functions give
\begin{eqnarray}
 \delta m_{H_1}^2 &=& \mathcal{O}(M_R^{-1}),\nonumber \\
 \delta m_{H_2}^2 &=& \frac{Y_\nu^2}{8\pi^2}\left[m_{\tilde{N}}^2 + {\rm Re}(A_\nu) B_N \right] + \mathcal{O}(M_R^{-1}), \nonumber \\
 \delta B &=& \frac{Y_\nu^2}{16\pi^2} B_N + \mathcal{O}(M_R^{-1}),
\end{eqnarray}
which are the same as Eq.(\ref{eq:threshold}).

Next, we discuss how the numerical value of the parameter $\mu$  can be affected by threshold corrections for the Higgs bilinear terms
in the radiative electroweak symmetry breaking scenario
\cite{Inoue:1982ej,Inoue:1982pi,Inoue:1983pp,Ibanez:1982fr,AlvarezGaume:1981wy}. 
In the calculation, we assume that gaugino masses, scalar masses and
A-terms are universal at the GUT scale.

\section{mu term and radiative electroweak symmetry breaking}
%In this section, we discuss how the value of the parameter $\mu$ can be affected by threshold corrections to the electroweak symmetry breaking. --> I think this sentence is repeated by the above sentence (last sentence in the previous section)
As we have shown, the soft breaking parameter for the Higgs mass
$m_{H_2}^2$ in the MSSM at the seesaw scale $M_R$ is determined by not only $\bar{m}_{H_2}^2(M_R^2)$ calculated via RGEs in the SUSY seesaw model but also additional contribution  due to the loops mediated by light and heavy sneutrinos  in the seesaw model at the scale $M_R$ .
From Eq.(\ref{eq:threshold}), the shift of $m_{H_2}^2$ from $\bar{m}_{H_2}^2$ at the scale $M_R$ is approximately given as,
\begin{eqnarray}
 \delta m_{H_2}^2 &\approx& \frac{Y_\nu^2}{8\pi^2} {\rm Re}.(A_\nu B_N) \nnn
&\approx& 1.6 \times 10^5 ({\rm GeV})^2 \left(\frac{Y_\nu}{0.5}\right)^2 \left(\frac{{\rm Re}A_\nu}{100 {\rm GeV}}\right) \left(\frac{B_N}{500 {\rm TeV}}\right) .
\label{eq:shift_th}
\end{eqnarray}
Therefore the soft breaking parameter $B_N$ of the order of $500\  {\rm TeV}$ may significantly affect  $m_{H_2}^2$ at the scale $M_R$.
This observation in turn indicates that the shift of $m_{H_2}^2$ at the scale  $M_R$ affects electroweak symmetry breaking in the MSSM
when we take the MSSM as an effective theory of SUSY type I seesaw model at low energy scale.

Let us discuss how electroweak symmetry breaking can be affected by the parameter $B_N$.
In the MSSM, radiative breaking of electroweak symmetry can occur when SSB parameters for Higgs sectors satisfy the following relation:
\begin{eqnarray}
\frac{1}{2} m_Z^2 = - |\mu|^2 + \frac{m_{H_1}^2(m_Z^2) - m_{H_2}^2(m_Z^2) \tan^2\beta}{\tan^2\beta-1} . \label{eq:ewsbcond}
\end{eqnarray}
In the limit of large $\tan\beta$, this relation becomes
\begin{eqnarray}
 \frac{1}{2} m_Z^2 \approx - |\mu|^2 - m_{H_2}^2(m_Z^2) . \label{appEWSB}
\end{eqnarray}
Therefore we see that the value of $\mu$ and $m_{H_2}^2$ are directly related.
In order to satisfy this condition, $m_{H_2}^2$ has to be negative at the scale $m_Z$.
In the radiative electroweak symmetry breaking scenario, $m_{H_2}^2$ is generally taken to be positive at high energy scale, but
it receives quite large radiative corrections due to heavy stop mass and large top quark Yukawa couplings between high and low energy scales, which drive $m_{H_2}^2$ negative so that electroweak symmetry can break at low energy scale.
At the scale above $M_R$, soft breaking masses and couplings are subject to
the RGEs of the SUSY seesaw model. The RGE for $m_{H_2}^2$ in the SUSY seesaw model is given by \cite{seesaw_rge, seesaw_rge2}
\begin{eqnarray}
\frac{d m_{H_2}^2}{dt} = \frac{2}{16\pi^2}\left[
-\frac{3}{5} g_1^2 M_1^2 - 3g_2^2 M_2^2 + 3 Y_t^2 X_t + Y_\nu^2 X_n
\right] , \label{eq:rge_seesaw}
\end{eqnarray}
where $t=\ln\frac{Q}{Q_0}$, $X_t = m_{\tilde{Q}_3}^2 + m_{\tilde{t}_R}^2 + m_{H_2}^2 + |A_t|^2$ and $X_n = m_{\tilde{L}}^2 + m_{\tilde{N}}^2 + m_{H_2}^2 + |A_\nu|^2$.
Here, $M_1$ and $M_2$ denote the bino mass and the wino mass, respectively.
The last term comes from the presence of RH neutrino superfields and other terms are the same as those in the MSSM.

It is expected that the RGE for $m_{H_2}^2$ can be significantly affected by the Yukawa coupling of neutrino sector $Y_{\nu}$
when it is quite large. We can estimate the deviation of the $m_{H_2}^2$ from that without neutrino sector by integrating out 
eq.({\ref{eq:rge_seesaw}}) explicitly. The deviation at the scale $M_R$ is approximately given as,
\begin{eqnarray}
 \delta_{log} m_{H_2}^2 \approx \frac{Y_\nu^2}{8\pi^2} (3m_0^2 + A_0^2) \ln \frac{M_R}{M_{X}} ,
\end{eqnarray}
where $m_0$ and $A_0$ are the universal values for scalar masses and A-terms respectively.
For $M_R = 6\times 10^{13}$ GeV and $M_X \approx 2\times10^{16}$ GeV, this contribution can be written approximately 
as,
\begin{eqnarray}
 \delta_{log} m_{H_2}^2 &\approx& -5.5 \times 10^4 ({\rm GeV})^2 \left(\frac{Y_\nu}{0.5}\right)^2 \left(\frac{m_0}{1 {\rm TeV}}\right)^2 . 
\label{eq:shift_log}
\end{eqnarray}
As we can see from eq.(\ref{eq:shift_th}), $\delta_{log} m_{H_2}^2$ is easily dominated by the threshold correction when $B_N$ is large.

%%%%
Without threshold corrections, the weak scale value of $m_{H_2}^2$ becomes more negative than that of minimal supergravity (mSUGRA) case. This affects the condition for
electroweak symmetry breaking and the allowed regions for the observed relic density of dark matter\cite{Calibbi, Barger}.
Especially, the allowed region where $|\mu|$ is small, is changed significantly. Universal scalar mass 
at the GUT scale, $m_0$ is larger than that of mSUGRA.
%so-called focus point region where $|\mu|$ is small, hardly exists\cite{Calibbi}.
However with inclusion of the threshold corrections, $m_0$ can be smaller than that of mSUGRA when $B_N$ is large.

\begin{figure}[htbp]
\includegraphics[width=12cm]{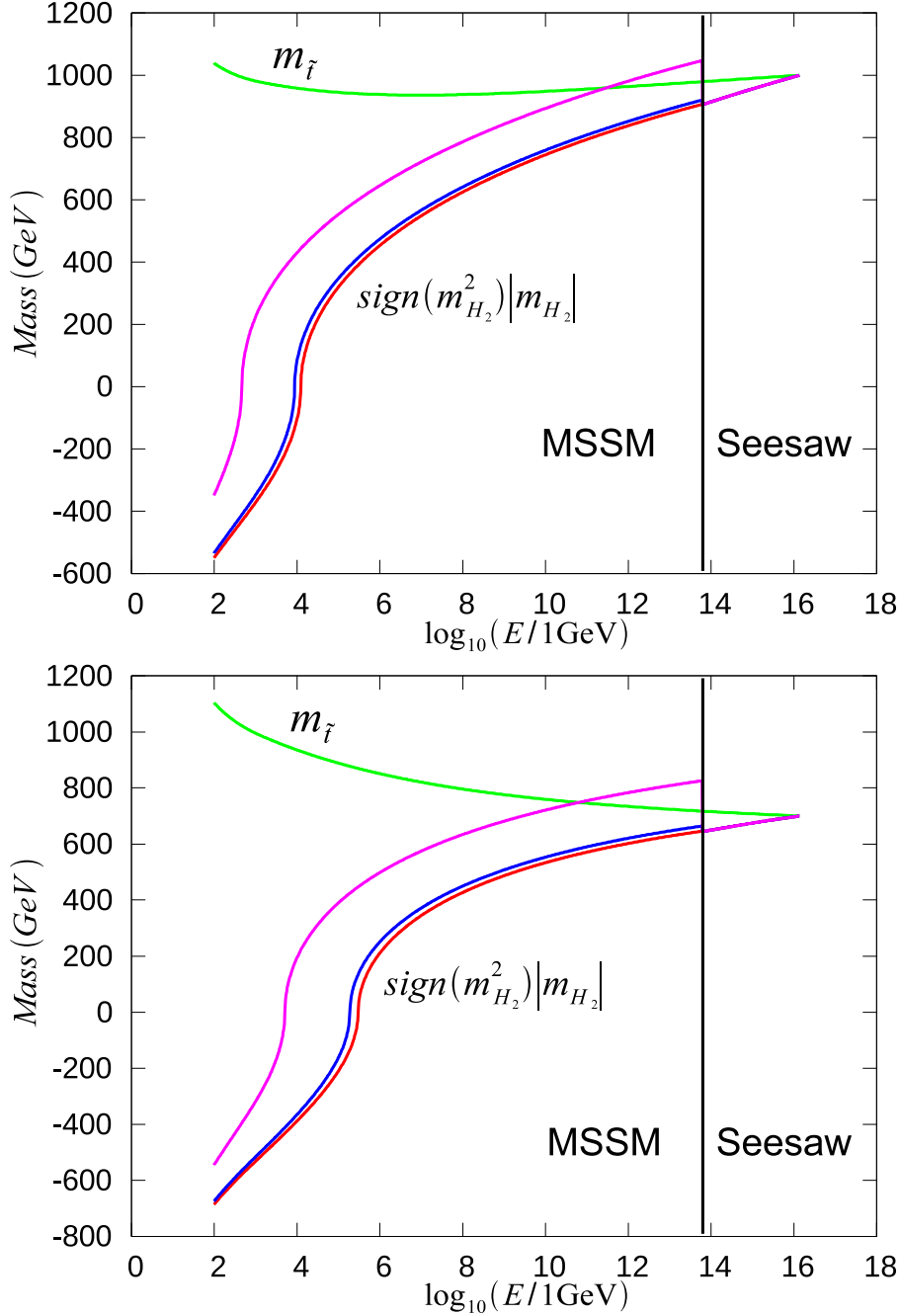}
\caption{The renormalization group evolutions of soft scalar masses for up-type Higgs and stops are shown.
The calculation is performed by taking $m_0, m_{1/2}, A_0, \tan\beta$ to be
$1 {\rm TeV}, 400 {\rm GeV}, 300 {\rm GeV}, 10$, respectively in the upper panel and
$700 {\rm GeV}, 500 {\rm GeV}, 300 {\rm GeV}, 20$, respectively in the lower panel.
We take neutrino masses $m_\nu$ and $M_R$ to be 0.1eV and $6\times 10^{13}$GeV, respectively, in both figures 
so that $Y_\nu$ and $Y_t$ become the same order of magnitude.
The pink, blue and red curves correspond to the predictions of $sign(m^2_{H_2})|m_{H_2}|$
for $B_N=500,  50$ and $5$ TeV, respectively.
The green curves correspond to the MSSM prediction of $m_{\tilde{t}}$.
}
\label{fig:rgeevo}
\end{figure}

Figure \ref{fig:rgeevo} shows the RG evolution of $m_{H_2}^2$ and $m_{\tilde{t}}^2$ with the energy scale. Here, $m_{\tilde{t}}$ is defined as
$m_{\tilde{t}}^2=m_{\tilde{Q}} m_{\tilde{t}_R}$.
We assume that soft breaking masses, gaugino masses and A-terms are universal at the GUT scale($\approx 2 \times 10^{16} {\rm GeV}$).
The calculations are performed with ISASUGRA code which is included in ISAJET package \cite{isajet}.
The input values used in the calculations are given in the caption and neutrino masses $m_\nu$ and $M_R$  are taken to be 0.1 eV and $6\times 10^{13}$GeV, respectively, in both panels so that $Y_\nu$ and $Y_t$ become the same order of magnitude.
The pink, blue and red curves correspond to the predictions of $sign(m^2_{H_2})|m_{H_2}|$ including threshold corrections
for $B_N=500, 50$ and $5$ TeV, respectively.
The green curves show how the predictions of $m_{\tilde{t}}^2$ evolve from the GUT scale to the electroweak scale.
%%%%%%%%added
When $B_N = 50\, {\rm TeV}$, $A_\nu \sim 300\, {\rm GeV}$ and $m_0 \sim 1\,{\rm TeV}$,
the threshold correction and the running effects from the neutrino Yukawa sector are almost canceled, 
i.e. $\delta m_{H_2}^2 + \delta_{log} m_{H_2}^2 \sim 0$.
Therefore the blue lines below the scale $M_R$ behave as if there are no effects from neutrino Yukawa sector.
%%%%%%%%%%%%%%
As we can see from Fig. \ref{fig:rgeevo},
the value of $m_{H_2}^2$ at the scale $m_Z$ obtained in the SUSY seesaw model is significantly deviated from that obtained
in the MSSM for given input values of $m_0, m_{1/2}, A_0, \tan \beta$ and $B_N=500 {\rm TeV}$, 
whereas such a deviation disappears for $B_N \lesssim 5 {\rm TeV}$.
%%%%%%%%%%%%%%%compare

In the case without threshold corrections, the running of the $m_{H_2}^2$ in mSUGRA with RH neutrino superfield (mSUGRA+RHN) is discussed 
in Refs. \cite{Barger,seesaw_DM5}. The weak scale values of $\sqrt{|m_{H_u}^2}|$ tend to be larger than those in mSUGRA scenario. 
The difference between mSUGRA and mSUGRA+RHN is up to a few hundred GeV, when $m_0 \gtrsim 1.5\, {\rm TeV}$ and $Y_\nu \gtrsim Y_t$. 
On the other hand, our results show that 
the threshold correction increases $m_{H_u}(Q^2=M_R^2)$ by several hundred GeV and therefore the weak scale values of $\sqrt{|m_{H_u}^2}|$
can be smaller than those in mSUGRA scenario when $B_N$ is large.

\begin{figure}[htbp]
\includegraphics[width=12cm]{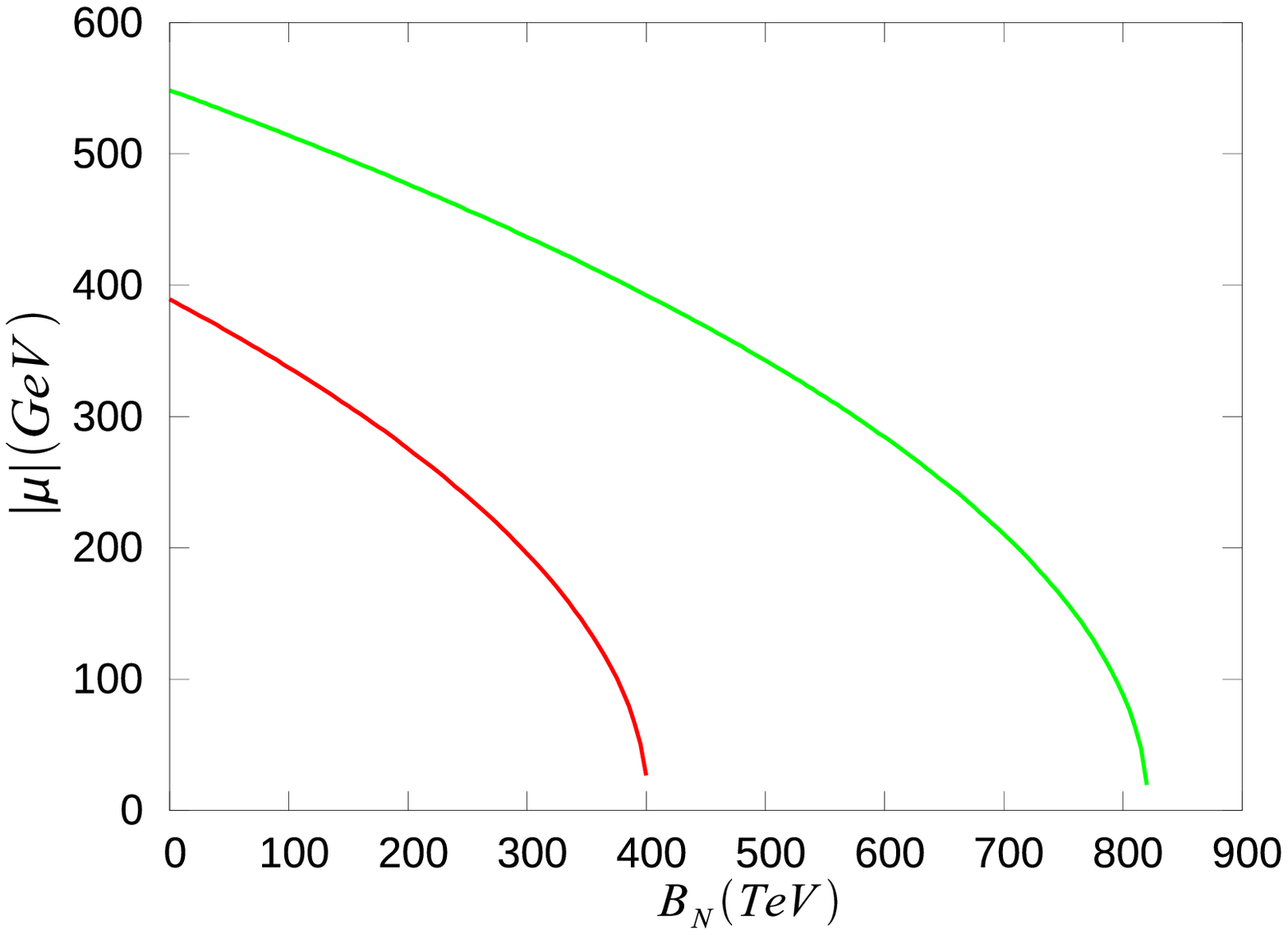}
\caption{The values of $|\mu|$ are plotted as a function of $B_N$. The lower red line is obtained
for $m_0=1 {\rm TeV}, m_{1/2}=400 {\rm GeV}, A_0=300 {\rm GeV}$ and $\tan\beta=10$, and the upper green line for
$m_0=700 {\rm GeV}, m_{1/2}=500 {\rm GeV}, A_0=300 {\rm GeV}$ and $\tan\beta=20$.
We take the same values of  $m_\nu$ and $M_R$ as in Fig. 1.
}
\label{fig:mu}
\end{figure}

The significant deviation of $m_{H_2}^2$ at the scale $m_Z$ in turn leads to a significant change in $|\mu|$ through the stationary condition, Eq. ({\ref{eq:ewsbcond}}).
In Fig. {\ref{fig:mu}}, we present how  $|\mu(M_Z)|$ depends on the value of $B_N$.
As the value of $B_N$ increases, $|\mu|$ becomes smaller, due to the threshold corrections to $m_{H_2}^2(M_R)$.

It is worthwhile to notice that the size of the mass parameter $\mu$ characterizes the property of neutralino dark matter.
Since $\mu$ is the Higgsino mass term, changing $\mu$ may affect the composition
of the neutralino dark matter.
This indicates that relic abundance of the dark matter is affected by $B_N$, especially on the condition that gaugino masses are universal
at the GUT scale.

\section{Bino-Higgsino Dark matter}
In this section, we show that the lightest SUSY particle (LSP) is a bino-Higgsino mixture state when the size of parameter
$B_N$ is of the order of several hundred TeV, and the result of the WMAP observation can be accounted for well.
Here, we assume that soft scalar masses, gaugino masses and $A$ terms are universal at the GUT scale.
We consider the lightest neutralino as a dark matter candidate.

The neutralinos are the physical states which are composed of the bino, wino
 and two Higgsinos.
The neutralino mass matrix in the $\tilde{B}{\tiny -}\tilde{W}{\tiny -}\tilde{H_1}{\tiny -}\tilde{H_2}$ basis
is given by,
\begin{eqnarray}
 \mathcal{M}_{\chi} = \left(
\begin{array}{cccc}
 M_1 & 0 & -m_Z \cos\beta \sin\theta_W& m_Z \sin\beta \sin\theta_W \\
 0 & M_2 & m_Z \cos\beta \cos\theta_W& -m_Z \sin\beta \cos\theta_W \\
 -m_Z \cos\beta \sin\theta_W & m_Z \cos\beta \cos\theta_W & 0 & -\mu \\
 m_Z \sin\beta \sin\theta_W & -m_Z \sin\beta \cos\theta_W & -\mu & 0\\
\end{array}
\right) \ ,
\end{eqnarray}
where $M_1$ and $M_2$ are the bino and wino masses, respectively, and $\theta_W$ is the Weinberg angle.
%$\mu$ denotes the Higgsino mass parameter, determined by theEq.(\ref{eq:ewsbcond}).  ---> I think that this sentence is useless, mu is already defined in sec. II.
This matrix is diagonalized by the unitary matrix $N$,
\begin{eqnarray}
 \mt{M}_{\chi}^{diag} = N^* \mt{M}_{\chi} N^{-1} .
\end{eqnarray}
In terms of $N$, the lightest neutralino $\chi^0$ is expressed as a mixture of the gauginos and the Higgsinos:
\begin{eqnarray}
 \chi^0 = N_{11} \tl{B} + N_{12} \tl{W} + N_{13} \tl{H}_1 + N_{14} \tl{H}_2 .
\end{eqnarray}

Since we assumed a universal value for gaugino masses at the GUT scale, gaugino masses
$M_i$ are related to gauge couplings $g_i$ as follows;
\begin{eqnarray}
 \frac{M_i(Q)}{M(\Lambda_{GUT})} = \frac{g_i^2(Q)}{g^2(\Lambda_{GUT})}, \label{eq:gmass}
\end{eqnarray}
and this relation is easily derived  from renormalization group equations for gauginos,
\begin{eqnarray}
 \frac{dM_i}{dt} = \frac{2}{16\pi^2} b_i g_i^2 M_i ,
\end{eqnarray}
where $b_i$ are coefficients of beta-functions for $g_i$.
From Eq.(\ref{eq:gmass}), the bino mass $M_1$ is written in terms of the wino mass $M_2$:
\begin{eqnarray}
 M_1 = \frac{5}{3} \tan^2\theta_W M_2 \approx 0.5 M_2,
\end{eqnarray}
at the scale $m_Z$.

The relic density of a cold dark matter, $\Omega_{\rm CDM} h^2$, is determined by WMAP observation \cite{wmap} and its value is given by
\begin{eqnarray}
 \Omega_{\rm CDM} h^2 = 0.1131 \pm 0.0034 .
\end{eqnarray}
For $|\mu| \gg M_2$, the dark matter is bino-like, whereas for $|\mu| \ll M_2$ the dark matter is Higgsino-like .
In general, a bino-like dark matter leads to a large relic abundance of a dark matter, which can not accommodate the result from WMAP observation.
%
%This is because the couplings of the vertices which include bino-like dark matter are much smaller than the couplings for Higgsino-like and wino-like dark matters, so the annihilation cross section for pure bino-like dark matter is quite small.
This is because couplings for bino are smaller than those for Higgsino and wino.
When the value of $|\mu|$ decreases, the Higgsino fraction defined by $|N_{13}|^2 + |N_{14}|^2$ increases, which  leads to larger
annihilation cross sections for Higgsino-like dark matter. Therefore we can fit the right amount of relic abundance derived from the result of WMAP observation with dark matter candidate composed of a bino-Higgsino mixture.

\begin{figure}[htbp]
\includegraphics[width=12cm]{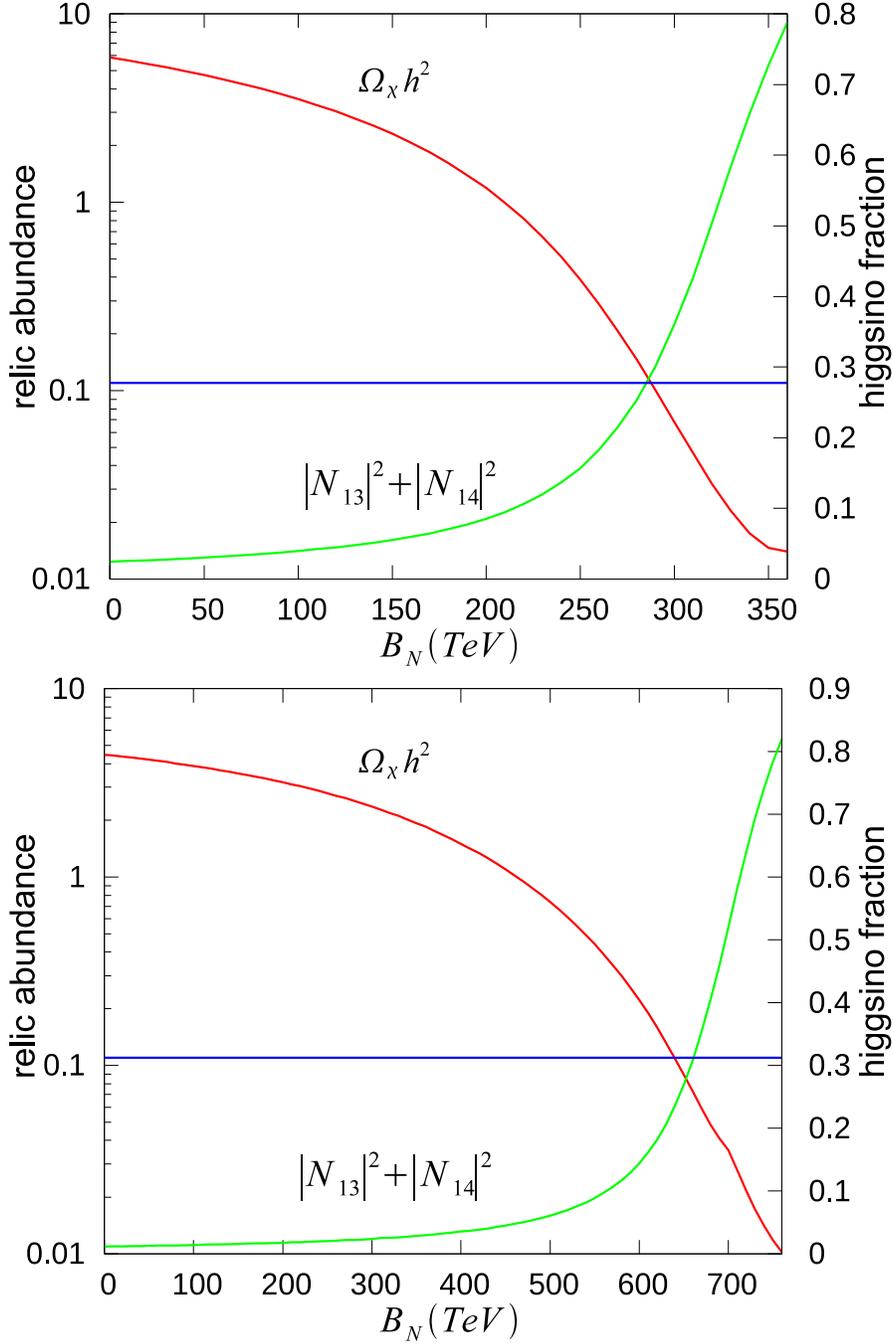}
\caption{The relic abundances of the lightest neutralino (red curves) and corresponding Higgsino contributions (green curves) are drawn as a function of $B_N$. We
take $m_0, m_{1/2}, A_0, \tan\beta$ to be $1 {\rm TeV}, 400 {\rm GeV}, 300 {\rm GeV}, 10$, respectively in the upper panel and to be $700 {\rm GeV}, 500 {\rm GeV}, 300 {\rm GeV}, 20$, respectively in the lower panel. $\mu$ is positive, and the values of $m_\nu$  and $M_R$ are taken to be the same as in Fig. 1.
The blue lines correspond to the value of the relic abundance obtained from WMAP observation.}
\label{fig:relic}
\end{figure}

As we can see from Eq.(\ref{eq:ewsbcond}), since the value of $|\mu(m_Z)|^2$ becomes smaller as $B_N(M_R)$ increases.
Larger value of $B_N(M_R)$  leads to larger Higgsino fraction, which makes the relic abundance of a dark matter decreased.
Fig. \ref{fig:relic} presents the predictions of relic abundance of the lightest neutralino 
and  corresponding contributions of Higgsino components as a function of $B_N$. Our numerical calculation is
performed by using micrOMEGAs 2.2 code \cite{micromegas1, micromegas2}.
The blue line represents the value of the relic abundance obtained from WMAP observation.
In this figure, we can see that
as $B_N(M_R)$ increases, Higgsino fractions get larger, which makes relic abundances smaller.
From our numerical analysis, it turned out that the right amount of the relic abundance of the dark matter could be explained by
 taking the parameter $B_N$ to be of the order of several hundred TeV which makes Higgsino fractions large.
%%%%%%%%%%%%%%%%%%%%%%%%%%%%%%%%%%%%%%%%%%%
The allowed regions of parameter space for the observed relic density of the dark matter are most conveniently shown
in ($m_{1/2}$,$m_0$) plane.
In mSUGRA+RHN scenario without threshold correction, the allowed regions
are given in Refs.\cite{Barger,seesaw_DM5}.
One of the regions corresponding to the small $\mu$ is located along the region where electroweak symmetry breaking can not take place.
This region corresponds to $m_0 \gtrsim 1.3 {\rm TeV}$. The values of $m_0$ depend on
%$M_R$ and $Y_\nu$ significantly.
the renormalization group running effect from neutrino Yukawa sector and it decreases the low energy value of $m_{H_2}^2$.
When this effect becomes larger, we need to choose larger $m_0$ as the GUT boundary condition.
% llowd region where $\mu$ is small, shifts 
%this region shift to larger values of $m_0$ and $m_{1/2}$.
 With inclusion of the threshold correction to $m_{H_2}^2$, however, the consequences change.
In our scenario, as shown in Fig.{\ref{fig:relic}}, we can take $m_0$ as small as $700 {\rm GeV}$, since the threshold correction
 is added to $m_{H_2}^2$ at the scale $M_R$.
%this region can be smaller values of $m_0$, such as $m_0 = 700\, {\rm GeV}$ and $m_0 = 1\, {\rm TeV}$ due to the threshold correction.
%%%%%%%%%%%%%%%%%%%%%%%%%%%%%%%%%%%%%%%%%%%%%%%%%%
% so that the required value of the annihilation cross section could be obtained.
Therefore, we conclude that the allowed regions where the observed relic density is explained by the bino-Higgsino dark matter
%of parameters space for $m_0, m_{1/2}, A_0$ and $\tan\beta$ 
are very different from those of 
mSUGRA and mSUGRA+RHN scenario.

\section{Conclusion and discussion}
We have investigated the effective low energy Higgs potential of the SUSY type I seesaw model.
We found that  Higgs bilinear terms got threshold corrections at the scale below $M_R$, due to  heavy singlet sneutrino loops.
These threshold corrections are proportional to B-term of heavy singlet sneutrino $B_N$. Therefore if $B_N$ is large enough,
the mass parameters of Higgs bilinear terms are significantly shifted at the scale $M_R$, which in turn leads to a shift of
the parameter $|\mu|$ and reduction of the fine-tuning  between the Higgsino mass parameter $\mu$ and SSB mass for up-type Higgs at the electroweak scale. 
We presented how the parameter $\mu$ depends on $B_N$.
We have shown that dark matter becomes a mixture of bino and Higgsino for $B_N$ of the order of several hundreds TeV and the observed relic abundance can be consistently explained by the bino-Higgsino dark matter.
It turned out that the allowed region of parameter space constrained by the relic abundance of dark matter in this model
is very different from the
MSSM without seesaw under the assumption that SSB terms are universal at the GUT scale,
mainly because of the threshold corrections to $m_{H_2}^2$. %Because of inclusion of the large threshold corrections,
Our results are also different from those of conventional mSUGRA with type I seesaw which does not include the threshold corrections to
$m_{H_u}^2$.

Naturalness problem for such a large value of $B_N$ is beyond the scope of this work.
Since the size of $B_N$ of order of several hundreds TeV is much larger than the scale of soft breaking parameters,
the origin of $B_N$ must be different from those of other SUSY breaking parameters.
$U(1)_{B-L}$ extension of the MSSM might provide the origin of large $B_N$.
It would be interesting if such a large value of $B_N$ can be naturally possible.
%If such a large value of $B_N$ can be naturally possible, there are significant effects in low energy physics through threshold corrections to
%soft breaking terms for Higgs sector. Therefore we have to consider the effects from $B_N$ correctly.

\section{acknowledgement}
We would like to thank participants
of Summer Institute 2009, phenomenology, for discussion
where the preliminary
result of the work was presented by N. Y.
We also would like to thank Lorenzo Calibbi for pointing out the
importance of the renormalization group running effects
and the related references.
The work of S.K.Kang was supported in part by the Korea Research Foundation(KRF) grant funded by the Korea government(MEST) (2009-0069755), and
the work of T. M. was supported by KAKENHI,
Grant-in-Aid for Scientific Research on Priority Areas,
Mass Origin and SuperSymmetric Physics (No.16028213),
and New Development for Flavor Physics (No.19034008 and
No.20039008), MEXT, Japan.
\appendix
\section{CP violation and phase convention}
Here, we discuss the CP violation of SUSY type I seesaw model
and identify the independent phases by choosing a phase
convention. One can assign
the R charge $0$ to the Higgs superfields
$\hat{H}_1$ and $\hat{H}_2$, and $1$ to the lepton superfields $\hat{L}$ and $\hat{N}^c$.
Under the R transformation and the phase redefinition
of the superfields $\hat{L}, \hat{H_1},\hat{H_2}, \hat{N}^c $,
the super potential is transformed as,
\bea
W && \rightarrow e^{2 i \theta_R} \left(
-Y_\nu \exp(i(\theta_{N^c}+ \theta_L + \theta_2))
\hat{N}^c \hat{L} \cdot \hat{H_2}
-\frac{M_R}{2} \exp(2 i \theta_{N^c}) \hat{N^c}\hat{N^c} + \right. \nn \\
&& \left. \mu
\exp(i(\theta_1+\theta_2-2 \theta_R)) \hat{H_1} \cdot \hat{H_2}
) \right).
\eea
Therefore one can remove the phases of the parameters
$Y_{\nu}$, $\mu$, $M_R$ in $W$ by choosing the phases of the superfields as follows,
\bea
&& \theta_{N^c}=-\frac{1}{2}\arg{M_R}, \nn \\
&& \theta_1+\theta_2-2 \theta_R =-\arg \mu, \nn \\
&& \theta_L +\theta_2+ \theta_{N^c} =-\arg{Y_{\nu}},
\label{eq:phase}
%&&& \theta_Q+\theta_2+\theta_{t^c}=-\arg{Y_t}
\eea
The trilinear couplings of the
soft breaking terms transform in the same way as
the super potential, so one can not remove those phases.
For the soft breaking parameters of the bilinear form, one can take one of them to be real.
We then rotate the phase of $B_N$ away by choosing the phase parameter
of R transformation as follows,
\bea
\theta_R=-\frac{1}{2} \arg(B_N).
\eea
In Eq.(\ref{eq:phase}),
we still have the freedom of choosing the
phase of $\theta_2$.
Here, we choose the phase $\theta_2$ so that vacuum expectation value
of $H_2$ becomes real
\bea
\theta_2=-\arg(v_2).
\eea
To summarize we choose the phases as,
\bea
&& \theta_{N^c}=-\frac{1}{2}\arg{M_R} \nn \\
&& \theta_1=-\arg \mu+\arg(v_2)-\arg{B_N} \nn \\
&& \theta_L=\arg(v_2)+\frac{1}{2} \arg{M_R}-\arg{Y_{\nu}}.
%&& \theta_Q+\theta_{t^c}=-\arg{Y_t}+\arg(v_2). \nn \\
\eea
With this phase convention, the soft breaking terms are
written as,
\bea
{\cal L}_{soft}&=&\left(|A_{\nu}| |Y_{\nu}|\tilde{N^\ast}
e^{i(\arg \frac{A_\nu}{B_N})}
+ h.c. \right)
\nn \\
&+&2 |\mu| |B| {\rm Re}.( e^{i(\arg \frac{B}{B_N})} H_1 \cdot H_2) -
\frac{|M_R|}{2} |B_N|
 \tilde{N}^{\ast} \tilde{N}^{\ast}
\nn \\
&-& m_{\tilde{L}}^2 |\tilde{L}|^2- m_{\tilde{N}}^2 |\tilde{N}|^2,
\eea
and two independent irremovable CP violating phases are presented as,
\bea
B&=&|B| e^{i\arg\frac{B}{B_N}}, \nn \\
A_{\nu}&=& |A_\nu| e^{i \arg\frac{A_{\nu}}{B_N}}.
%A_{t}&=& |A_t| e^{i \arg\frac{A_{t}}{B_N}}.
\eea

\section{derivation of the effective potential}
In this appendix, we derive the effective potential of Higgs fields in SUSY type I seesaw model.
The contribution to the effective
potential for Higgs fields from the loops mediated by neutrino superfields is written as,
\bea
V_{\rm eff}(v_1, v_2)=\int \frac{Q^{4-d} d^d k}{(2 \pi)^d i}
\frac{1}{2}
\left( \ln \rm \det({M_s}^2-k^2)
-\ln \rm \det(M_F-k\hspace{-.67em}/) \right), \label{eq:ep1}
\eea
where $M_F$ is the mass matrix of one of the neutrino sector and $M_s^2$
is the 4 by 4 mass-squared matrix of sneutrino sector given by
\bea
{M_s}^2=\left(\begin{array}{cccc}
  (m_{\tilde L}^2+m_D^2) & 0 &
 \hat{A}_\nu^\ast m_D  & |M_R m_D| \\
  0 & ( m_{\tilde L}^2+m_D^2) & |M_R m_D| &
\hat{A}_\nu m_D \\
\hat{A}_\nu m_D &  |m_D M_R| & |M_R|^2 + m_{\tilde N}^2 & |B_N M_R|  \\
|m_D M_R| & \hat{A}_\nu^{\ast} m_D
& |B_N^{\ast} M_R| & |M_R|^2 +m_{\tilde N}^2
\end{array} \right),
\eea
where
\bea
m_D&=&\frac{Y_\nu v_2}{\sqrt{2}} \nn \\
\hat{A}_{\nu}&=&A_{\nu}-\frac{v_1^{\ast}}{v_2}\mu \nn \\
A_{\nu}&=&|A_{\nu}|e^{i{\arg\frac{A_{\nu}}{B_N}}}.
\eea
The effects of CP violation appear through the parameter $\hat{A}_{\nu}$.
We compute the following quantity,
\bea
\ln \rm \det({M_s}^2-k^2)=Tr Ln ({M_s}^2-k^2).
\eea
To compute the scalar contribution,
we diagonalize $M_s^2$ approximately and treat
$A$ term as perturbation.
We first split $M_s^2$ as
\bea
{M_s}^2=M_0^2+\Delta_A \ ,
\eea
where
\bea
M_0^2 &=&
\left(\begin{array}{cccc}
  (m_{\tilde L}^2+m_D^2) & 0 &
 0 & |M_R m_D| \\
  {\bf 0} & ( m_{\tilde L}^2+m_D^2)  & |M_R m_D| &
0 \\
0 &  |m_D M_R| & |M_R|^2 + m_{\tilde N}^2+m_D^2& |B_N M_R|  \\
|m_D M_R| & 0
& |B_N^{\ast} M_R| & |M_R|^2 +m_{\tilde N}^2+m_D^2
\end{array} \right) ,
\label{eq:Orth}
\end{eqnarray}
and
\begin{eqnarray}
\Delta_A &=&
\left(\begin{array}{cccc}
  0 & {\bf 0} &
 \hat{A}_\nu^\ast m_D  & 0 \\
  {\bf 0} & 0 & 0 & \hat{A}_{\nu} m_D \\
\hat{A}_\nu m_D &  0 & 0 & 0 \\
0 & \hat{A}_\nu^{\ast} m_D
& 0 & 0
\end{array} \right) .
\end{eqnarray}
One can find the orthogonal matrix $O$ which diagonalizes
$M_{0}^2$. Using this matrix, ${M_s}^2$ is transformed as
\bea
O Ms^2 O^T= {\rm diag}(m_1^2, m_2^2, m_3^2, m_4^2) +  O \Delta_A O^T.
\eea
Here, $m_1,m_2$ are the mass of lighter sneutrinos and
$m_3,m_4$ are those of heavier sneutrinos given by
\bea
m_1^2&=& \frac{M_R^2+m_{\tilde N}^2+ 2 m_D^2+B_N M_R +m_{\tilde L}^2}{2}
- \frac{1}{2} \sqrt{(M_R^2+m_{\tilde N}^2+ B_N M_R
-m_{\tilde L}^2)^2+4 m_D^2 M_R^2} , \nn  \\
m_2^2&=& \frac{M_R^2+m_{\tilde N}^2+ 2 m_D^2-B_N M_R +m_{\tilde L}^2 }{2}
- \frac{1}{2} \sqrt{(M_R^2+m_{\tilde N}^2-B_N M_R
-m_{\tilde L}^2)^2+4 m_D^2 M_R^2} , \nn \\
m_3^2&=& \frac{M_R^2+m_{\tilde N}^2+ 2 m_D^2-B_N M_R +m_{\tilde L}^2}{2}
+\frac{1}{2} \sqrt{(M_R^2+m_{\tilde N}^2- B_N M_R
-m_{\tilde L}^2)^2+4 m_D^2 M_R^2}
, \nn \\
m_4^2&=&\frac{M_R^2+m_{\tilde N}^2+2 m_D^2+B_N M_R
+m_{\tilde L}^2}{2}
+\frac{1}{2} \sqrt{(M_R^2+m_{\tilde N}^2+B_N M_R
-m_{\tilde L}^2)^2+4 m_D^2 M_R^2} . \nn \\
\eea
These sneutrino masses should be compared with the neutrino masses written as,
\bea
m_H^2&=&\frac{M_R^2}{2}+m_D^2+ \frac{\sqrt{M_R^4+ 4 m_D^2 M_R^2}}{2} , \nn \\
m_L^2&=&\frac{M_R^2}{2}+m_D^2-\frac{\sqrt{M_R^4+ 4 m_D^2 M_R^2}}{2} .
\eea
Using Eq.(\ref{eq:ep1}) and the mass eigenvalues, one can find effective potential as follows,
\bea
V_{\rm eff}=V^{(0)}_{\rm eff}
+\int \frac{d^d k }{(2 \pi)^d i}
\frac{1}{2}\left(
 + \sum_{n=1}^{\infty} \frac{(-1)^{n-1}}{n}
{\Tr} \left(\frac{1}{m^2-k^2} O \Delta_A O^T \right)^n \right) \ ,
\eea
where
\bea
V^{(0)}_{\rm eff}&=&\frac{1}{2}
\int \frac{d^d k Q^{4-d}}{(2 \pi)^d i}\left(
\sum_{i=1}^4 \log(m_i^2-k^2)- 2 \log(m_H^2-k^2)-2 \log(m_L^2-k^2)
\right) \nn \\
&=&\frac{1}{64 \pi^2}
C_{UV}
\left(2 (m_H^4+m_L^4)-\sum_{i=1}^4 m_i^4 \right) \nn \\
&+&\frac{1}{64 \pi^2} \left(
\sum_{i=1}^4 m_i^4 (\log \frac{m_i^2}{Q^2}-\frac{3}{2})
-2 m_H^4(\log \frac{m_H^2}{Q^2}-\frac{3}{2})
-2 m_L^4(\log \frac{m_L^2}{Q^2}-\frac{3}{2}) \right) , \nnn
\eea
where $C_{UV}=\frac{1}{\epsilon}-\gamma+\log 4 \pi$
and $Q$ is the renormalization scale.
The renormalization point dependent finite part of the effective potential
$V^{(0)}_{\rm eff}$
is given as,
\bea
V^{(0)}(Q^2)=
\frac{1}{64 \pi^2} \left(
\sum_{i=1}^4 m_i^4 (\log \frac{m_i^2}{Q^2}-\frac{3}{2})
-2 m_H^4(\log \frac{m_H^2}{Q^2}-\frac{3}{2})
-2 m_L^4(\log \frac{m_L^2}{Q^2}-\frac{3}{2}) \right).
\eea
We note that $V^{(0)}$ depends on the Higgs vacuum expectation
value through $m_D^2$ where
$m_D=\frac{Y_{\nu} v_2}{\sqrt{2}}$.
To obtain the contribution to
the Higgs mass term $m_{H_2}^2 H_2^{\dagger}
H_2$, one can differentiate the effective potential with respect to $m_D^2$,
while keeping the terms which remain
non zero in large limit of $M_R$,
\bea
\frac{\partial V^{(0)}}{\partial m_D^2}&\simeq&
\frac{1}{64 \pi^2}\left(
(\log \frac{M_R^2}{Q^2}-C_{UV}-1)
(2 m_3^2 \frac{\partial m_3^2}{ \partial m_D^2}
+2 m_4^2 \frac{\partial m_4^2}{ \partial m_D^2}-4 m_H^2
\frac{\partial m_H^2}{ \partial m_D^2}) \right. \nn \\
&+& \left. 2(m_3^2 \log \frac{m_3^2}{M_R^2}
\frac{\partial m_3^2}{ \partial m_D^2}
+m_4^2 \log \frac{m_4^2}{M_R^2}
\frac{\partial m_4^2}{ \partial m_D^2}
-2 m_H^2 \log \frac{m_H^2}{M_R^2}
\frac{\partial m_H^2}{ \partial m_D^2}) \right) \nn \\
&\simeq& \frac{1}{16 \pi^2}\left( (\log \frac{M_R^2}{Q^2})
(m_{\tilde L}^2+ m_{\tilde N}^2)+2 m_{\tilde N}^2 \right)
-\frac{1}{16 \pi^2}(C_{UV}+1)
(m_{\tilde L}^2+ m_{\tilde N}^2),
\label{first}
\eea
The terms which are proportional to the derivative
of
the lighter mass also vanish in large limit of $M_R$, because $m_1^2 \sim m_2^2 \simeq m_{\tilde L}^2 \simeq \frac{m_D^4}{M_R^2}$ and the derivatives
with respect to $m_D^2$ are suppressed as $\frac{B_N}{M_R}$
and  $\frac{m_D^2}{M_R^2}$, respectively.
From Eq. (\ref{first}), one can read off the coefficient of the
Higgs mass term $H_2^{\dagger} H_2$.
The contribution to the Higgs mass term including the
counter term is given as,
\bea
V_{\rm eff}^{(0)}(Q^2)&=&V_{\rm eff}^{(0)}+V_c^{(0)} \nn \\
            &=&\frac{Y_\nu^2}{16 \pi^2}(H_2^{\dagger} H_2)
\left(\log \frac{M_R^2}{Q^2}(m_{\tilde L}^2+m_{\tilde N}^2)+2 m_{\tilde N}^2 \right)
\eea
where the counter term is given as,
\bea
V_c^{(0)}=\frac{{Y_\nu}^2}{16 \pi^2} (C_{UV}+1) (m_{\tilde L}^2
+m_{\tilde N}^2) H_2^{\dagger} H_2.
\eea

Next we compute the corrections to $V^{(0)}$ due to the
$A_\nu$ terms up to the second order of $\Delta_A$, because
they give the non-vanishing contribution to
the effective potential in large limit of $M_R$.
To compute the corrections, one needs to derive the
orthogonal
matrix $O$ in Eq.(\ref{eq:Orth}).
To diagonalize $M_0^2$ follows two steps. First,
we diagonalize $M^2_0$ with the help of orthogonal matrices $O_L$ and $O_H$ as follows,
\bea
M_0^{\prime 2}
&=& \left(
\begin{array}{cc}
O_L & 0 \\
0  & O_H \end{array} \right) M_0^2 \left(\begin{array}{cc}
O_L^T & 0 \\
0  & O_H^T \end{array} \right) \nn \\
&& = \left(\begin{array}{cccc}
m_{\tilde L}^2+m_D^2& 0 & 0 & m_D M_R \\
0 & m_{\tilde L}^2+m_D^2 & m_D M_R & 0 \\
0 & m_D M_R & M_R^2+m_{\tilde N}^2+m_D^2-B_N M_R & 0 \\
m_D M_R & 0 & 0 & M_R^2+m_{\tilde N}^2 +m_D^2+B_N M_R
\end{array} \right) , \nnn
\eea
where $O_L$ and $O_H$
are given as
\bea
O_L=O_H^T= \frac{1}{\sqrt{2}} \left(\begin{array}{cc}
1 & 1 \\
-1 & 1
\end{array}
\right) .
\eea
We note the
degenerate diagonal masses of the
heavy sneutrinos are split after the rotation.
The mass squared matrix $M_0^{\prime 2}$ has the separated
two by two parts as sub-matrices.  Each of them
has the form of the seesaw type. Thus, the
mass matrix $M_0^{\prime}$ can be diagonalized as,
\bea
&&\left(\begin{array}{cccc}
m_1^2 & 0& 0& 0 \\
0 & m_2^2 & 0 & 0 \\
0 & 0 & m_3^2& 0 \\
0 & 0 & 0 & m_4^2
\end{array} \right)= \nn \\
&&
\left(\begin{array}{cccc}
\cos \theta_+ & 0 & 0 & -\sin \theta_+ \\
0 & \cos \theta_- & -\sin \theta_- & 0 \\
0 & \sin \theta_-  & \cos \theta_- & 0 \\
\sin \theta_+ & 0 & 0 & \cos \theta_+
\end{array} \right) M_0^{\prime 2}
\left(\begin{array}{cccc}
\cos \theta_+ & 0 & 0 & \sin \theta_+ \\
0 & \cos \theta_- & \sin \theta_- & 0 \\
0 & -\sin \theta_-  & \cos \theta_- & 0 \\
-\sin \theta_+ & 0 & 0 & \cos \theta_+
\end{array} \right).
\eea
Then the orthogonal matrix $O$ is  given as
\bea
O=\left(\begin{array}{cccc}
\cos \theta_+ & 0 & 0 & -\sin \theta_+ \\
0 & \cos \theta_- & -\sin \theta_- & 0 \\
0 & \sin \theta_-  & \cos \theta_- & 0 \\
\sin \theta_+ & 0 & 0 & \cos \theta_+
\end{array} \right) \times \left( \begin{array}{cc}
O_L & 0 \\
0 & O_H \end{array}
\right).
\eea
\def\dif{\theta_- + \theta_+}
Using the above form of orthogonal matrix $O$,  $O \Delta_A O^T $ is given as
\bea
O \Delta_A O^T &=&m_D {\rm Re}(\hat{A}_\nu)
\left(
\begin{array}{cccc}
-\sin 2 \theta_+ & 0 & 0 & \cos 2 \theta_+ \\
0 & \sin 2 \theta_- & - \cos 2 \theta_- & 0 \\
0 & -\cos 2 \theta_- & -\sin 2 \theta_- & 0 \\
\cos 2 \theta_+ & 0 & 0  & \sin 2 \theta_+
\end{array} \right) + \nn \\
&& i m_D {\rm Im}
(\hat{A}_\nu)
\left(
\begin{array}{cccc}
0 & \sin (\dif) & -\cos (\dif) & 0 \\
-\sin \dif & 0 & 0 & \cos (\dif) \\
\cos (\dif) & 0 & 0 & \sin (\dif) \\
0 & -\cos (\dif) & -\sin (\dif) & 0
\end{array}
\right). \nnn
\eea
We then obtain the corrections to the effective potential
at the first order of $\Delta_A$ given as
\bea
\delta V_{\rm eff}^{(1)}&=&\frac{1}{2}
\left(\sum_{i=1}^4 \int \frac{d^d k}{(2 \pi)^d i}
\frac{(O \hat{A}_{\nu}
O^T)_{ii}}{m_i^2-k^2} \right)\nn \\
&=& -\frac{Re.\hat{A}_{\nu} m_D}{32 \pi^2} \times \nn \\
&& \left(m_4^2 \sin 2 \theta_+(C_{UV}+1-\ln \frac{m_4^2}{Q^2})
 -m_1^2 \sin 2 \theta_+ (C_{UV}+1 -\ln \frac{m_1^2}{Q^2}) \right. \nn \\
&-&  \left. m_3^2 \sin 2 \theta_-(C_{UV}+1-\ln \frac{m_3^2}{Q^2})
+m_2^2 \sin 2 \theta_- (C_{UV}+1-\ln \frac{m_2^2}{Q^2})
\right).
\eea
Now, let us show how the divergences are canceled so that the correction is finite.
To do this, we use the relation
\bea
(m_4^2-m_1^2) \sin 2 \theta_+=(m_3^2-m_2^2) \sin 2 \theta_- .
\eea
Then, the corrections to the effective potential become
\bea
\delta V_{\rm eff}^{(1)}&=&\frac{m_D Re. \hat{A}_\nu}{32 \pi^2}
\left(m_1^2 \sin 2 \theta_+ \ln \frac{m_3 m_4}{m_1^2}-m_2^2
\sin 2 \theta_- \ln \frac{m_3 m_4}{m_2^2} \right.\nn \\
&+& \left.\frac{m_3^2 \sin 2 \theta_- + m_4^2 \sin 2 \theta_+ }{2}
\ln \frac{m_4^2}{m_3^2} \right) \nn \\
&\simeq& \frac{m_D Re.(\hat{A}_\nu)} {32 \pi^2}
(m_4^2-m_3^2) (\theta_+ + \theta_-) \nn \\
&\simeq & \frac{m_D^2}{8 \pi^2} Re(\hat{A}_{\nu} B_N) \nn \\
&\simeq & \frac{Y_{\nu}^2}{8 \pi^2}
Re.(A_{\nu} B_N \frac{v_2^2}{2}- \mu B_N
\frac{v_1^{\ast}{v_2}}{2}) \nn \\
&\simeq&\frac{Y_{\nu}^2}{8 \pi^2}
\left( Re.(A_{\nu} B_{N}) H_2^{\dagger} H_2-
\mu B_N Re.(H_1 \cdot H_2) \right),
\eea
where we have used the relation which is valid in large limit of $M_R$,
$\theta_{\pm} \sim \frac{m_D}{M_R}$ and
$m_4^2-m_3^2=2 M_R B_N$.
The correction  at the second order of $\Delta_{A_\nu}$ term is given
as
\bea
\delta V_{\rm eff}^{(2)}=-\frac{1}{4} \int \frac{d^d k}{(2 \pi)^d i}
\frac{1}{m_i^2-k^2} (O \Delta_A O^T)_{ij}\frac{1}{m_j^2-k^2}
(O \Delta_A O^T)_{ji}.
\eea
The term which is not suppressed by $\frac{1}{M_R^n}$ is given as,
\bea
\delta V_{eff}^{(2)}&=&-\frac{1}{16 \pi^2}
(C_{UV}+1-\ln\frac{M_R^2}{Q^2})
m_D^2 |\hat{A}_{\nu}|^2 \nn \\
&=& -\frac{{Y_\nu}^2}{16 \pi^2}(C_{UV}+1-\ln \frac{M_R^2}{Q^2})
\left(|A_{\nu}|^2 H_2^{\dagger} \cdot H_2
-2 {\rm Re} (A_\nu \mu H_1 \cdot H_2)+
\mu^2 H_1^{\dagger} \cdot H_1 \right).\nn \\
\eea
The divergences are canceled by adding the counter term,
\bea
V_c^{(2)}
&=& \frac{{Y_\nu}^2}{16 \pi^2}(C_{UV}+1)
\left(|A_{\nu}|^2 H_2^{\dagger} \cdot H_2
-2 {\rm Re} (A_\nu \mu H_1 \cdot H_2)+
\mu^2 H_1^{\dagger} \cdot H_1 \right).
\eea
The effective potential at one loop level is finally written as
\bea
V_{\rm eff}^{\rm 1 loop}
&=&  \left(|\mu|^2 + m_{H_1}^2(Q^2) \right)  H_1^{\dagger}H_1
+\left(|\mu|^2 + m_{H_2}^2(Q^2) \right)  H_2^{\dagger}H_2
-2 {\rm Re.} (B(Q^2) \mu H_1 \cdot {H}_2)
\nn \\
&+&  \left(\mu^2 \frac{Y_\nu^2}{16 \pi^2}
\log \frac{M_R^2}{Q^2} \right) H_1^{\dagger} H_1 \nn \\
&+&  \frac{Y_{\nu}^2}{16 \pi^2}
\left(
\log \frac{M_R^2}{Q^2} (m_{\tilde L}^2+m_{\tilde N}^2
+|A_\nu^2|)
+2 m_{\tilde N}^2 + 2 {\rm Re.}(A_\nu B_N)
\right)H_2^{\dagger}H_2 \nn \\
&-&2{\rm Re.} \left( \frac{Y_\nu^2}{16 \pi^2}
(B_N + A_\nu \log \frac{M_R^2}{Q^2}) \mu
H_1 \cdot H_2 \right) - \mathcal{L}_D, \nn
\eea
where $\mathcal{L}_D$ is the D-term contribution.
To complete the renormalization of the effective potential, we consider the relation between
the renormalized mass parameters and the bare ones.
We first note that the bilinear part of the Higgs
sector including the tree and the counter terms
in the present model can be derived
from the following Lagrangian,
\bea
{\cal L}&=&Z_1 \hat{H}_1^{\dagger} \hat{H}_1|_D+Z_2
\hat{H}_2^{\dagger} \hat{H}_2|_D+
\mu \hat{H}_1 \cdot \hat{H}_2|_F
+ h.c. \nn \\
&-& (m_{H_1}^2(Q^2)+ \delta m_{H_1}^2) H_1^{\dagger} H_1 -
(m_{H_2}^2(Q^2)+ \delta m^2_{H_2})  H_2^{\dagger} H_2
\nn \\
&+& 2 {\rm Re}.\left(
(B(Q^2)+\delta B)\mu H_1 \cdot H_2 \right).
\eea
After integrating out $F$ terms of the superfields,
one obtains,
\bea
{\cal L}&=&Z_1 \partial_\mu  H_1^{\dagger} \partial^{\mu} H_1
+Z_2 \partial_\mu H_2^{\dagger} \partial^{\mu} H_2 \nn \\
&-& \frac{|\mu|^2}{Z_2} H_1^\dagger H_1-\frac{|\mu|^2}{Z_1}
H_2^{\dagger} H_2 + 2 {\rm Re}.\left(
(B(Q^2)+\delta B)\mu H_1 \cdot H_2 \right)\nn \\
&-& (m_{H_1}^2(Q^2)+ \delta m_{H_1}^2)
H_1^{\dagger} H_1-(m_{H_2}^2(Q^2)+ \delta m_{H_2}^2)
H_2^{\dagger} H_2.
\label{eq:l}
\eea
We define bare superfields and  bare parameter  $\mu$  as
$
\hat{H_i^0}=\sqrt{Z_i}\hat{H_i}$, $(i=1,2)$ and
$\mu_0 \sqrt{Z_1} \sqrt{Z_2}=\mu$, respectively.
One can write the Lagrangian in terms of the bare fields as,
\bea
{\cal L}&=& \partial_\mu  H_1^{0 \dagger} \partial^{\mu} H_1^0
+\partial_\mu H_2^{0 \dagger} \partial^{\mu} H_2^0
-|\mu_0|^2 H_1^{0 \dagger} H_1^0-|\mu_0|
H_2^{0 \dagger} H_2^0 + 2 Re(B_0 \mu_0 H_1^0 \cdot H_2^0)\nn \\
&-&\frac{(m_{H_1}^2(Q^2)+ \delta m_{H_1}^2)}{Z_1}
H_1^{0 \dagger} H_1^0-\frac{m_{H_2}^2(Q^2)+ \delta m_{H_2}^2}{Z_2}
H_2^{0 \dagger} H_2^0.
\eea
Then one can define the bare mass parameters
as,
\bea
m_{0 H_1}^{2}Z_1&=&m_{H_1}^2(Q^2)+\delta m_{H_1}^2, \nn \\
m_{0 H_2}^{2}Z_2&=&m_{H_2}^2(Q^2)+\delta m_{H_2}^2, \nn \\
B_0&=& B(Q^2)+ \delta B.
\eea
Eq.(\ref{eq:l}) leads to the following counter terms for the bilinear
parts of the Higgs potential,
\bea
V_c&=&(\delta m^2_{H_1}+(Z_2^{-1}-1)|\mu|^2) H_1^\dagger H_1+
+ (\delta m^2_{H_2}+(Z_1^{-1}-1) |\mu|^2)  H_2^{\dagger} H_2 \nn \\
&-& 2 {\rm Re} (\delta B \mu H_1 \cdot H_2).
\eea
Comparing $V_c$ with the sum of the counter terms
$V_c^{(0)}+V_c^{(2)}$ given by
\bea
V_c^{(0)}+V_c^{(2)}&=&\frac{Y_\nu^2}{16 \pi^2}(C_{UV}+1)
(|A_\nu|^2+m_{\tilde N}^2+m_{\tilde L}^2) H_2^\dagger H_2
\nn \\
&+&\frac{Y_{\nu}^2}{16 \pi^2}(C_{UV}+1) \mu^2 H_1^\dagger H_1 \nn \\
&-2 &\frac{Y_\mu^2}{16 \pi^2}(C_{UV}+1)
{\rm Re}(A_\nu \mu H_1 \cdot H_2),
\eea
we obtain the following relations,
\bea
\delta m^2_{H_1}+(Z_2^{-1}-1) \mu^2&=&\frac{Y_{\nu}^2}{16 \pi^2}(C_{UV}+1)
\mu^2,
\nn \\
\delta m^2_{H_2}+(Z_1^{-1}-1) \mu^2&=& \frac{Y_\nu^2}{16 \pi^2}(C_{UV}+1)
(|A_\nu|^2+m_{\tilde N}^2+m_{\tilde L}^2), \nn \\
\delta B&=&\frac{Y_\mu^2}{16 \pi^2}(C_{UV}+1) A_\nu.
\eea
Using the results of the wave function renormalization,
\bea
Z_1&=&1, \nn \\
Z_2&=&1-\frac{Y_\nu^2}{16 \pi^2}C_{UV},
\eea
we obtain
\bea
\delta m^2_{H_1}&=&\frac{Y_\nu^2}{16 \pi^2} \mu^2,\\ \nn \\
\delta m^2_{H_2}&=&\frac{Y_{\nu}^2}{16 \pi^2}(
|A_\nu|^2+m_{\tilde N}^2+m_{\tilde L}^2) (C_{UV}+1).
\eea
Finally, we find the following relations between the renormalized
parameters and the bare ones,
\bea
m_{H_1}^2(Q^2)&=& m_{0 H_1}^2 -\frac{Y_\nu^2}{16 \pi^2}
\mu^2 , \nn \\
m_{H_2}^2(Q^2)&=& m_{0 H_2}^2 Z_2 -\frac{Y_\nu^2}{16 \pi^2}(|A_\nu^2|+
m_{\tilde N}^2+m_{\tilde L}^2) (C_{UV}+1),
 \nn \\
B(Q^2)&=& B_0 - \frac{Y_\nu^2}{16 \pi^2} A_\nu
(C_{UV}+1), \nn \\
\mu(Q^2)&=&\mu_0 \sqrt{Z_2}.
\eea

\end{document}